\documentclass[a4paper,11pt]{article}
\usepackage{amsmath,psfrag,bbm,epsfig}
\numberwithin{equation}{section}
\numberwithin{figure}{section}
\newcommand{\be}{\begin{equation}}
\newcommand{\ee}{\end{equation}}
\newcommand{\sembrack}[1]{[\![#1]\!]}

\title{
\vskip -70pt
\begin{flushright}
\normalsize{DAMTP-2004-43}
\end{flushright}
\vskip 25pt
\bf Area Regge Calculus and \\ Discontinuous Metrics}
\vspace{1.4cm}
\author{Chris $\mathrm{Wainwright^{1,}}$\thanks{e-mail address: C.J.Wainwright@damtp.cam.ac.uk} \hspace{1pt}
and Ruth M. $\mathrm{Williams^{1,2,}}${\thanks{e-mail address:
    R.M.Williams@damtp.cam.ac.uk}} \vspace{0.2cm}
\\
\small{\textsl{$\mathrm{{}^1}$DAMTP, Wilberforce Road, Cambridge CB3 0WA, England}} \\ 
\small{\textsl{$\mathrm{{}^2}$Girton College, Cambridge CB3 0JG, England.}}}
\date{\today}
\begin{document}
\maketitle

\begin{abstract}
Taking the triangle areas as independent variables in the theory of Regge
calculus can lead to ambiguities in the edge lengths, which can be
interpreted as discontinuities in the metric. We construct
solutions to area Regge calculus using a triangulated lattice and find
that on a spacelike hypersurface no such discontinuity can arise. On a
null hypersurface however, we can have such a situation and the resulting
metric can be interpreted as a so-called refractive wave.
\end{abstract}

\section{Introduction}

The Ponzano-Regge \cite{ponzanoregge} model of quantum gravity in
three dimensions begins with a labelling of the 1-simplices in a
simplicial complex with irreducible representations of $SU(2)$. Using
these labellings a \emph{state sum} is constructed, which is the
discrete analogue of the partition function of three dimensional
quantum gravity. The sum over labellings of the 1-simplices in the
state sum corresponds to an integration over edge lengths in the path
integral. The action appearing in the exponential of the path integral
is the Regge action \cite{regge:orig}, defined in terms of these edge lengths.

In four dimensions the labelling of the 1-simplices does not seem to
be sufficient \cite{discrete}. For example in the Barrett-Crane model \cite{barrettcrane} we label
the \emph{2-simplices} with balanced irreducible representations of
$SU(2)\times SU(2)$. The resulting state sum then takes the form of
the path integral for the four dimensional Regge action, yet now the
areas of the triangles appear to be taking the role of the independent
variables.

This is then our motivation to study Regge calculus with triangle
areas, rather than edge lengths, as the independent variables. This
idea was originally suggested by Rovelli \cite{rovelli} in connection
with loop quantum gravity.

In Section 2 we will introduce the theory of area Regge calculus and
discuss some if its potential consequences such as discontinuous
metrics. In Section 3 we will construct full solutions of area Regge
calculus to investigate the possibility of discontinuities along
spacelike and null hypersurfaces. We then, in Section 4, need to introduce the theory of tensor distributions which allows us,
in Section 5, to draw some interesting comparisons between area Regge
calculus and the theory of gravity.

\section{Area Regge Calculus}

In area Regge calculus we begin with the Regge action on a 4-dimensional simplicial complex \cite{regge:orig},

\be
I=\sum_{\Delta}A_{\Delta}\epsilon_{\Delta}.
\ee

\noindent However, instead of taking the lengths of the 1-simplices to be the independent variables we consider the areas of the 2-simplices, i.e. the $A_\Delta$, to be independent. Variation of the action with respect to the $A_\Delta$ results in the field equations $\epsilon_\Delta=0$, which states that the deficit angle at each triangle is zero. If this were standard Regge calculus, a zero deficit condition would imply the space were locally flat. However, the situation here is less straightforward, as explained in \cite{noteonareavar}, which we shall now summarize.

Consider a single 4-simplex, which contains 10 edges and 10 triangles. The edge lengths uniquely specify the geometry of the simplex and hence determine the areas. However the reverse is not true; two 4-simplices can be constructed having the same triangle areas, yet different edge lengths. This problem can be overcome if we consider our 4-simplices to be in some sense close to a regular 4-simplex, analogous to choosing the principal value of a multi-valued function such as $\sin^{-1}x$.

Now suppose we have two 4-simplices meeting at a tetrahedron. We now have a total of 16 triangles and 14 edges, hence a problem of over-determinism. Also note the shared tetrahedron has 4 triangles and 6 edges, hence an under-determined geometry. We can thus envisage a situation where calculating the edge lengths of the tetrahedron from the areas of one 4-simplex results in a different answer to if they were calculated from the other 4-simplex. We could interpret such a situation as a discontinuity of the metric.

We shall proceed by investigating under what circumstances such discontinuities can arise in solutions to area Regge calculus.

\section{Hypersurface Discontinuities}

We have seen in Section 2 how discontinuities might arise in area Regge calculus; however this was only for two adjoining simplices. We would like to be able to construct a full solution to area Regge calculus containing such discontinuities.

The simplest situation would be to restrict the discontinuity to some hypersurface $\Sigma$, separating $M$ into $M^\pm$. We triangulate $M$ such that we also triangulate $\Sigma$. Taking the areas of the triangles as the independent variables, we have the field equations $\epsilon_\Delta=0$, i.e. zero deficit angles.

To restrict the discontinuity to $\Sigma$ we demand the areas are chosen to give well defined edge lengths in $M^\pm$ respectively. Combined with the zero deficit condition we see that both $M^\pm$ are flat.

We simplify the situation further by assuming $\Sigma$ is embedded in
 $M^\pm$ with no extrinsic curvature; this takes care of the zero
 deficit condition by virtue of the fact that the interior dihedral
 angles at those triangles in $\Sigma$ will be $\pi$ in both $M^\pm$. We work with a triangulated
 lattice, each lattice site being identical to its neighbours except
 across $\Sigma$.

The procedure we will take is as follows.
\begin{itemize}
\item Begin with a flat 3-lattice in $\partial M^-$.
\item Transform the coordinates of the lattice points (thus keeping the lattice flat) in such a way that the areas of the triangles are unchanged. This gives the lattice on $\partial M^+$.
\item Extend the 3-lattices on $\partial M^\pm$ to 4-lattices on $M^\pm$ and since the triangle areas agree we can identify the lattices on $\partial M^\pm$ giving the full 4-dimensional solution.
\end{itemize}

We can look at the cases of $\Sigma$ spacelike and $\Sigma$ null separately.

\subsection{$\Sigma$ Spacelike}

\begin{figure}[htb]
\begin{center}
\psfrag{1}{$1$}
\psfrag{2}{$2$}
\psfrag{3}{$3$}
\psfrag{4}{$4$}
\psfrag{5}{$5$}
\psfrag{6}{$6$}
\psfrag{7}{$7$}
\epsfig{file=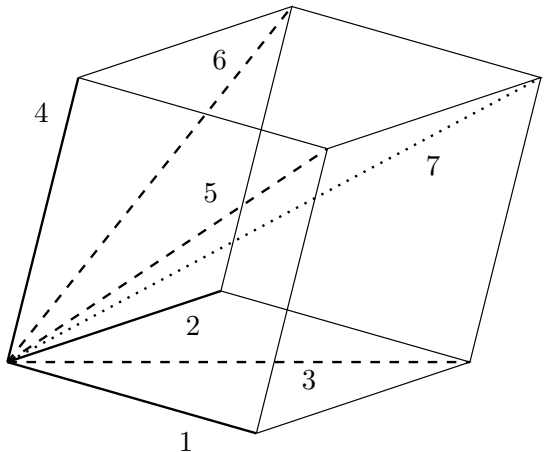}
\end{center}
\caption{Triangulated lattice edges}
\label{fig:edges}
\end{figure}

We work in coordinates $(t,x,y,z)$ with $\Sigma$ located at $t=0$. Our
flat 3-lattice in $\partial M^-$ is given by coordinates in
$(x,y,z)$. The metric in this space is $ds^2=dx^2+dy^2+dz^2$. We label the edges of the triangulated 3-lattice from 1 to
7 as shown in Figure \ref{fig:edges}, consistent with the binary
notation of \cite{qrc, qofrc}. Since
the hypersurface is flat the coordinates of edges 3, 5, 6 and 7 are
determined by those of edges 1, 2 and 4. 

Let $\boldsymbol{x}_i$ be the vector along the $i\mbox{'th}$ edge, then a
transformation of the lattice is determined by a transformation of
$\boldsymbol{x}_1$, $\boldsymbol{x}_2$ and $\boldsymbol{x}_4$. This
induces a linear transformation on $\mathbbm{R}^3$, which we write as a matrix $T\in GL(3)$.

Now since each lattice site is identical we only have 6 independent
triangle areas\footnote{Each lattice site has 12 triangles associated
  to it, however in the flat case each of these triangles has another
  which is always congruent to it, for example $A_{13}=A_{23}$, $A_{17}=A_{67}$, etc.}, these are $A_{13}, A_{15}, A_{17}, A_{26}, A_{27}$ and
$A_{37}$, where $A_{ij}$ is the area of the triangle which has $i$ and
$j$ as edges.
 
Under the transformation $\boldsymbol{x}_i \to \boldsymbol{x}'_i=T\boldsymbol{x}_i$ we require that these six areas remain
invariant. We can decompose $T$ in order to analyse which
transformations will satisfy this condition.

We proceed by first noting we can always split $T$ as $T=OP$, where
$O\in O(3)$ and $P$ is some positive definite symmetric matrix. Since
$P$ is symmetric we can further decompose it as $P=QDQ^T$, where $Q\in
O(3)$ and $D$ is diagonal with positive entries (since $P$ was
positive definite). Thus we have decomposed $T$ as

\be
T=OQDQ^T.
\ee

Since both $O$ and $Q$ are orthogonal they will not transform the
triangle areas. Thus we only require that $D$ does not transform the
triangle areas. So, let $D=\mbox{diag}(a,b,c)$ with $a,b,c>0$, and
consider a triangle based at the origin, with the other two vertices
at $(X_A,Y_A,Z_A)$ and $(X_B,Y_B,Z_B)$. This triangle has an area $A$ given by

\be
A^2=\frac{1}{4}\left(\left| \begin{array}{cc}
Y_A & Z_A \\
Y_B & Z_B  \\
\end{array} \right|^2+\left| \begin{array}{cc}
Z_A & X_A \\
Z_B & X_B  \\
\end{array} \right|^2+\left| \begin{array}{cc}
X_A & Y_A \\
X_B & Y_B  \\
\end{array} \right|^2\right),
\ee

\noindent and under a transformation by $D$ the area changes as

\be
A'^2=\frac{1}{4}\left(b^2c^2\left| \begin{array}{cc}
Y_A & Z_A \\
Y_B & Z_B  \\
\end{array} \right|^2+c^2a^2\left| \begin{array}{cc}
Z_A & X_A \\
Z_B & X_B  \\
\end{array} \right|^2+a^2b^2\left| \begin{array}{cc}
X_A & Y_A \\
X_B & Y_B  \\
\end{array} \right|^2\right).
\ee

\noindent Now, we require $A'=A$. Hence, if we let $\alpha=b^2c^2-1$,
$\beta=c^2a^2-1$ and $\gamma=a^2b^2-1$ we must have

\be\label{eqn:linear1}
\alpha\left| \begin{array}{cc}
Y_A & Z_A \\
Y_B & Z_B  \\
\end{array} \right|^2+\beta\left| \begin{array}{cc}
Z_A & X_A \\
Z_B & X_B  \\
\end{array} \right|^2+\gamma\left| \begin{array}{cc}
X_A & Y_A \\
X_B & Y_B  \\
\end{array} \right|^2=0.
\ee

\noindent This condition has to be satisfied for all 6 triangles,
giving 6 linear equations in $\alpha$, $\beta$ and $\gamma$. Letting
$\boldsymbol{v}^T=(\alpha,\beta,\gamma)$ we can write these in matrix
form,

\be\label{eqn:mat1}
G\boldsymbol{v}=0,
\ee

\noindent where $G$ is a $6\times 3$ matrix, each row consisting of the
squared determinants appearing in equation (\ref{eqn:linear1}) for each
of the six triangles.

We argue in appendix \ref{sec:rankg} that the
matrix $G$ will have rank 3. Hence the only solution to equation
(\ref{eqn:mat1}) will be $\boldsymbol{v}=0$. This in turn implies that
$a=b=c=1$ and thus only orthogonal transformations leave the 6
triangle areas independent.

\subsection{$\Sigma$ Null}

Now we wish to perform a similar analysis for the case when $\Sigma$
is null. We work in double null coordinates $(u,v,y,z)$ with $\Sigma$
located at $u=0$. Our flat three lattice is given by coordinates in
$(v,y,z)$. The metric in this space is simply $ds^2=dy^2+dz^2$, since
$v$ is a null direction. 

Again we transform the lattice by changing the three linearly
independent vectors defining it; this induces a $GL(3)$ transformation on
the $(v,y,z)$ space, which is given by the matrix $\tilde T$. We
proceed to decompose this transformation, as we did in the spacelike case, but
in a different fashion. Note now that areas of triangles do not depend
on the $v$ coordinates of their vertices, hence we begin by
writing $\tilde T$ as

\be
\left(\begin{array}{cc}
a & \boldsymbol{b}^T \\
\boldsymbol{c} & E
\end{array}\right)=
\left(\begin{array}{cc}
a-\boldsymbol{b}^TE^{-1}\boldsymbol{c} & \boldsymbol{b}^TE^{-1} \\
0 & \mathbbm{1}
\end{array}\right)\left(\begin{array}{cc}
1 & 0 \\
\boldsymbol{c} & \mathbbm{1}
\end{array}\right)\left(\begin{array}{cc}
1 & 0 \\
0 & E
\end{array}\right),
\ee

\noindent where $E\in GL(2)$ and $\boldsymbol{b}$, $\boldsymbol{c}$ 
are 2-component vectors. We can decompose $E$ as we did for $T$ in the
spacelike case, giving $E=\tilde O \tilde P=\tilde O\tilde Q\tilde
D\tilde Q^T$, with $\tilde O,\tilde Q\in O(2)$ and $\tilde D$ diagonal with
positive entries. Hence $\tilde T$ now takes the form

\be
\tilde T=\left(\begin{array}{cc}
\ast & \ast \\
0 & \mathbbm{1}
\end{array}\right)\left(\begin{array}{cc}
1 & 0 \\
\boldsymbol{c} & \mathbbm{1}
\end{array}\right)\left(\begin{array}{cc}
1 & 0 \\
0 & \tilde O\tilde Q
\end{array}\right)\left(\begin{array}{cc}
1 & 0 \\
0 & \tilde D
\end{array}\right)\left(\begin{array}{cc}
1 & 0 \\
0 & \tilde Q^T
\end{array}\right),
\ee

\noindent which can be rearranged to the form

\be\label{eqn:nulldecomp}
\tilde T=\left(\begin{array}{cc}
\ast & \ast \\
0 & \tilde O\tilde Q
\end{array}\right)\left(\begin{array}{cc}
1 & 0 \\
\boldsymbol{d} & \tilde D
\end{array}\right)\left(\begin{array}{cc}
1 & 0 \\
0 & \tilde Q^T
\end{array}\right),
\ee

\noindent where $\boldsymbol{d}=(\tilde O\tilde Q)^T\boldsymbol{c}$, and elements denoted by an asterisk ($\ast$) will not be relevant to our calculations.

We consider a triangle based at the origin with its other vertices at
$(V_A,Y_A,Z_A)$ and $(V_B,Y_B,Z_B)$. The area of the triangle is given by

\be
A=\frac{1}{2}\left|\begin{array}{cc}
Y_A & Z_A \\
Y_B & Z_B 
\end{array}\right|,
\ee

\noindent which does not depend on the $v$ coordinates of the vertices, as
stated earlier. We see that the first and third matrices in
the decomposition of equation (\ref{eqn:nulldecomp}) will not alter the
areas of the triangle, so all we require is that the second matrix does not either.

Let $\boldsymbol{d}^T=(c,d)$ and $\tilde D=\mbox{diag}(a,b)$ with
$a,b>0$, then what we require is that the triangle areas remain invariant
under $y'=ay+cv$ and $z'=bz+dv$. For our triangle's coordinates this gives the condition

\be\label{eqn:chris}
(ab-1)\left|\begin{array}{cc}
Y_A & Z_A \\
Y_B & Z_B
\end{array}\right|-
bc\left|\begin{array}{cc}
Z_A & V_A \\
Z_B & V_B
\end{array}\right|-
ad\left|\begin{array}{cc}
V_A & Y_A \\
V_B & Y_B
\end{array}\right|=0.
\ee 

Again we demand that the 6 triangle areas, $A_{13}, A_{15}, A_{17},
A_{26}, A_{27}$ and $A_{37}$, remain invariant under the transformation. Thus we
get 6 equations in our 3 unknowns $\tilde{\boldsymbol{v}}^T=(ab-1,
-bc, -ad)$, which we can write as

\be\label{eqn:nulllatvar}
\tilde G\tilde{\boldsymbol{v}}=0,
\ee

\noindent where $\tilde G$ is the $6\times 3$ matrix whose six rows contain
the determinants appearing in equation(\ref{eqn:chris}) for each of the six
triangles. We show in Appendix \ref{sec:ranktildeg} that the matrix $\tilde G$ has rank 3,
and hence the only solution to equation (\ref{eqn:nulllatvar}) is
$\tilde{\boldsymbol{v}}=0$. Thus we have $ab=1$ and $c=d=0$, which leads to
the most general form of $\tilde T$ leaving all six triangle areas
invariant as

\be\label{eqn:tnull}
\tilde T=\left(\begin{array}{cc}
\ast & \ast \\
0 & \tilde O\tilde Q
\end{array}\right)\left(\begin{array}{cc}
1 & 0 \\
0 & \tilde D
\end{array}\right)\left(\begin{array}{cc}
1 & 0 \\
0 & \tilde Q^T
\end{array}\right),
\ee

\noindent where again $\ast$ denotes elements which will not be relevant to us, and the matrix $\tilde D$ takes the form

\be\label{eqn:diag}
\tilde D=\left(\begin{array}{cc}
a & 0 \\
0 & a^{-1}
\end{array}\right),
\ee

\noindent with $a>0$.

\subsection{Metric discontinuities}

We can interpret these solutions to area Regge calculus as defining a metric on the space $M$ with a possible discontinuity at $\Sigma$. We began with a lattice in $\Sigma$, defined on a metric which we shall call $g$. The edge lengths of the lattice were defined in terms of this metric by $l^2=\boldsymbol{x}^Tg\boldsymbol{x}$. We then transformed the lattice with a linear transformation, which sent $\boldsymbol{x} \to T\boldsymbol{x}$. Thus the lengths of the edges were transformed as $l^2 \to l'^2=(T\boldsymbol{x})^Tg(T\boldsymbol{x})$. Equivalently, we can think of this as defining a new metric on $\Sigma$ given by

\be
g'=T^TgT.
\ee

\noindent Edge lengths are then calculated using the old coordinates $l'^2=\boldsymbol{x}^Tg'\boldsymbol{x}$.

For the case of $\Sigma$ spacelike, our initial metric was $g=\mathbbm{1}_3$, and we found that our transformation $T$ had to be orthogonal. Thus, the new metric is given by $g'=T^T\mathbbm{1}_3T=\mathbbm{1}_3$. So we see that in this case there is no difference in the metrics and the edge lengths will then be well defined across $\Sigma$.

For the case of $\Sigma$ null, our initial metric was of the form

\be
g=\left(\begin{array}{cc} 0 & 0 \\ 0 & \mathbbm{1}_2\end{array}\right),
\ee

\noindent and we found that $\tilde T$ must take the form given in equation (\ref{eqn:tnull}). Using this we can calculate the form of the new metric, which turns out to be

\be
g'=\left(\begin{array}{cc}
0 & 0 \\
0 & \tilde Q \tilde D^2 \tilde Q^T
\end{array}\right),
\ee

\noindent where $\tilde D$ takes the form given in equation (\ref{eqn:diag}). More explicitly, letting $\tilde Q$ be a rotation through angle $\theta$, we have

\be
g'=\left(\begin{array}{ccc}
0 & 0 & 0 \\
0 & a^2\cos^2\theta+a^{-2}\sin^2\theta & (a^2-a^{-2})\cos\theta\sin\theta \\
0 &(a^2-a^{-2})\cos\theta\sin\theta & a^2\sin^2\theta+a^{-2}\cos^2\theta
\end{array}\right).
\ee

This solution is precisely the metric of a refractive wave spacetime \cite{barrett:refract}.
Note that if we take $a$ to be close to unity, i.e. $a=1+\varepsilon
/2$, then to first order the difference in metrics on the hypersurface will take the form

\be\label{eqn:linrefract}
g'-g=\varepsilon\left(\begin{array}{ccc}
0 & 0 & 0 \\
0 & \cos2\theta& \sin2\theta \\
0 & \sin2\theta & -\cos2\theta
\end{array}\right),
\ee

\noindent and we can clearly see that the angle $\theta$ can be interpreted as the 
polarization of this wave, and the value $a$ as its strength.



\section{Tensor Distributions}

In order to handle discontinuous quantities properly we need to use
distributions. The standard idea of a distributions can be extended to
tensor valued distributions on arbitrary manifolds. There have been
many different approaches to this idea \cite{hartley,dray,geroch}; we follow here Geroch and
Traschen \cite{geroch}.

Just as we can define distributions on $\mathbbm{R}$ as linear maps
from a space of test functions to $\mathbbm{R}$, so we define tensor
distributions on a manifold $M$ as linear maps from a space of test
fields to $\mathbbm{R}$. The test fields in this case are tensor
densities of weight $+1$ with compact support, and we write the linear map as

\be
\hat T^{a\ldots}_{b\ldots}:\phi^{b\ldots}_{a\ldots} \to \langle
\hat T^{a\ldots}_{b\ldots},\phi^{b\ldots}_{a\ldots} \rangle ,
\ee 

\noindent which we need to satisfy certain continuity conditions, which are discussed more in \cite{geroch}.

Every smooth tensor field $S^{a\ldots}_{b\ldots}$ on $M$ defines a distribution by

\be\label{eqn:smoothdist}
\langle \hat S^{a\ldots}_{b\ldots},\phi^{b\ldots}_{a\ldots} \rangle = \int
S^{a\ldots}_{b\ldots}\phi^{b\ldots}_{a\ldots}\, dV.
\ee

\noindent The fact that we use densities, instead of just tensors as the
test fields means we do not need to introduce a volume element for
this integral, as is done in \cite{hartley,dray}.

We can define the derivative of distributions, which is consistent with that
for smooth tensors, by

\be
\langle \partial_c\hat T^{a\ldots}_{b\ldots},\phi^c{}^{b\ldots}_{a\ldots}
\rangle = -\langle \hat T^{a\ldots}_{b\ldots},\partial_c\phi^c{}^{b\ldots}_{a\ldots}\rangle.
\ee

\noindent If we introduce a smooth connection on $M$ then we have the
corresponding result

\be
\langle \nabla_c\hat T^{a\ldots}_{b\ldots},\phi^c{}^{b\ldots}_{a\ldots}
\rangle = -\langle \hat T^{a\ldots}_{b\ldots},\nabla_c\phi^c{}^{b\ldots}_{a\ldots}\rangle.
\ee

We can use the derivative of distributions to define the \emph{weak
  derivative} of locally integrable tensors\footnote{A tensor field
  $T^{a\ldots}_{b\ldots}$ is locally integrable if
  $T^{a\ldots}_{b\ldots}\phi^{b\ldots}_{a\ldots}$ is Lebesgue
  measurable and its integral converges for all test fields
  $\phi^{b\ldots}_{a\ldots}$. $T^{a\ldots}_{b\ldots}$ is locally
  square integrable if $T^{a\ldots}_{b\ldots}T^{c\ldots}_{d\ldots}$ is
  locally integrable}. A locally integrable tensor field
$T^{a\ldots}_{b\ldots}$ defines a distribution $\hat
T^{a\ldots}_{b\ldots}$, the same way as for a smooth tensor field in
  equation (\ref{eqn:smoothdist}). The weak derivative of $T^{a\ldots}_{b\ldots}$ is - if it exists - a locally integrable tensor field $R^{a\ldots}_{cb\ldots}$ such that $\hat R^{a\ldots}_{cb\ldots} = \nabla_c \hat T^{a\ldots}_{b\ldots}$.

The curvature tensor is nonlinear in the metric, and since
multiplication of distributions does not usually make sense, we need to
investigate which distributional metrics lead to a well-defined
curvature tensor.

Choosing an arbitrary smooth connection $\nabla_c$, with curvature
tensor $\rho^a{}_{bcd}$, we can write the curvature tensor for a
smooth metric $g_{ab}$ as

\be\label{eqn:curvtens}
R^a{}_{bcd}=\rho^a{}_{bcd}+2Q^a{}_{e[c}Q^e{}_{d]b}+2\nabla_{[c}Q^a{}_{d]b},
\ee

\noindent where

\be
Q^a{}_{bc}=\frac{1}{2}g^{ae}(2\nabla_{(b}g_{c)e}-\nabla_e g_{bc}).
\ee

\noindent Equation (\ref{eqn:curvtens}) continues to make sense if we relax the
smoothness conditions on $g_{ab}$. Geroch and Traschen show that a
distributional curvature tensor can still be defined for a so-called
\emph{regular metric}. A metric is said to be regular if (i) $g_{ab}$
and $g^{ab}$ exist everywhere and are locally bounded, and (ii) the
weak first derivative of $g_{ab}$ exists and is locally square
integrable.

Now consider a discontinuous metric. We split the manifold as
$M=M^+\cup M^-$ with $M^+\cap M^-=\Sigma$, where $\Sigma$ is a
codimension 1 smooth hypersurface. The metric is then given
by smooth tensor fields $g^\pm_{ab}$ defined on $M^\pm$
respectively. This metric clearly satisfies condition (i); to
calculate the weak first derivative we note that this metric defines
the distribution

\be
g_{ab}=\Theta^+g^+_{ab}+\Theta^-g^-_{ab},
\ee

\noindent where $\Theta^\pm$ are the step functions on $M^\pm$. The first derivative of this is

\be
\nabla_c g_{ab}=\Theta^+\nabla_c g^+_{ab}+\Theta^-\nabla_c
g^-_{ab}+\delta n_c \sembrack{g_{ab}},
\ee

\noindent where $\delta$ is the delta function on the boundary
$\partial M^-$, $n_c$ is the normal to this boundary, and
$\sembrack{g_{ab}}=g^+_{ab}|_{\partial M^-}-g^-_{ab}|_{\partial M^-}$. Since there
is no locally integrable tensor field which gives rise to the delta
function we see that if $\sembrack{g_{ab}}\ne 0$ then the weak
derivative of $g_{ab}$ does not exist, hence condition (ii) is not
met, and the metric is not regular. 

It is worth pointing out that distributional valued metrics appear
quite frequently in the literature, in particular with regard to
impulsive gravitational waves. Occasionally it is more convenient to
write the metric for such solutions in a coordinate system which is
not continuous. When this is done the metric may for instance contain
delta functions, making it unbounded and hence not
regular. However this is not a problem because the coordinate system
was considered discontinuous, and a transformation of
the coordinates back to a continuous coordinate system yields a
regular metric. For a nice discussion of this see
\cite{penrose}. Here, we are assuming the coordinate system to be
continuous, and as such the discontinuities cannot be transformed away
with such discontinuous coordinate transformations.

Another problem with a discontinuous metric is the ambiguity of
parallel transport. This is due to the fact there is no well defined
metric connection. The standard formula for the metric connection will
fail due to ill-defined multiplication of distributions, so the best
we can do is a so-called \emph{regularly discontinuous connection}
\cite{dray}. Here the connection components are continuous, and metric
compatible, in $M^\pm$ yet not across the boundary, which results in
the following rule for differentiation of the metric,

\be
\nabla_c g_{ab}=\delta n_c \sembrack{g_{ab}}.
\ee

Despite the lack of a natural rule for parallel transport, we can
still define metric geodesics. Since we are just extremizing the path
length this problem is well defined. The result is familiar from
optics, where we can describe the path of light through a medium with a
metric representing the time of travel along the path. Where the
refractive index of a medium changes abruptly we have a discontinuity
of this metric, and we know that the path of light is refracted at
this point. Exactly the same occurs here, with the geodesic refracted
at the boundary by the rule

\begin{eqnarray}
g^+(X^+,T)&=&g^-(X^-,T), \quad \forall T\in T_p\Sigma, \\
g^+(X^+,X^+)&=&g^-(X^-,X^-),
\end{eqnarray}

\noindent where $X^\pm$ is the tangent to the geodesic on either side
of the boundary.

It is interesting then that spaces which are solutions to area Regge
calculus may not have a well defined notion of parallel transport. 
However it seems we would be able to define geodesics in the
space.

\section{Comparisons with General Relativity}

We have shown in Section 4 that a discontinuity in the metric does not
make sense 
in general relativity, since the curvature tensors cannot be
defined. However, it is interesting to analyse the linearized Einstein
equations, since these allow distributional solutions of all
kinds. This will provide us with an interesting comparison with our area
Regge calculus solutions.

We shall work with the perturbation about flat space
$g_{ab}=\eta_{ab}+\varepsilon h_{ab}$ and write the linearized
equations as

\be\label{eqn:lineins}
G_{ef}^{abcd}\overline{h}_{ab,cd}=0,
\ee

\noindent where
$G_{ef}^{abcd}=\eta^{ac}\delta^b_e\delta^d_f+\eta^{ac}\delta^b_f\delta^d_e-\eta^{cd}\delta^a_e\delta^b_f-\eta_{ef}\eta^{ac}\eta^{bd}$
is a constant tensor and
$\overline{h}_{ab}=h_{ab}-\frac{1}{2}\eta_{ab}h$ is the trace reversed
perturbation. 

Considering $h_{ab}$ as a tensor distribution (from now on we
  will drop hats), we can find solutions to equation (\ref{eqn:lineins}) by its action on an arbitrary test tensor density $\phi^{ef}$.

\begin{eqnarray}
\langle G_{ef}^{abcd}\overline{h}_{ab,cd},\phi^{ef}\rangle &=&0 \\
\Rightarrow G_{ef}^{abcd}\langle\overline{h}_{ab},\phi^{ef}{}_{,cd}\rangle &=&0.
\end{eqnarray}

\noindent Substituting in the form of our desired discontinuous solution

\be
h_{ab}=\Theta^+p_{ab}+\Theta^-q_{ab},
\ee

\noindent we obtain the integral

\be
G_{ef}^{abcd}\int_{M^+}\overline{p}_{ab}\phi^{ef}{}_{,cd}\,dV + (q\mbox{-term}) = 0.
\ee

\noindent Integrating by parts twice gives

\be
G_{ef}^{abcd}\left[\,\int_{M^+}\overline{p}_{ab,cd}\phi^{ef}\,dV + \int_{\partial M^+}(\overline{p}_{ab}n_c\phi^{ef}{}_{,d}-\overline{p}_{ab,c}n_d\phi^{ef})\,dS\,\right] + (q\mbox{-term}) =0,
\ee

\noindent where $n^a$ is the normal to the hypersurface $\Sigma$. However
we should be careful, since the metric is discontinuous, the normal
vector will also be discontinuous across the hypersurface. In fact we
can define $n_\pm^a=n^a+\varepsilon m_\pm^a$ and then solve for
$m_\pm^a$. Fortunately, as the discontinuity in the normal direction is
${\cal O}(\varepsilon)$ it will not come into our calculations, but will
be important later.

Now $\phi^{ef}$ was arbitrary, hence choosing
$\mbox{supp}(\phi^{ef})\subset M^+$ only the first integral
contributes, hence we require the linearized Einsteins equations to be satisfied in $M^+$, i.e.

\be
G_{ef}^{abcd}\overline{p}_{ab,cd}=0.
\ee

\noindent Similarly $q_{ab}$ satisfies the linearized Einstein equations in $M^-$.

The more interesting result comes from the surface integral. To deal with this we must split the derivative into parallel and perpendicular parts, $\partial_c=\partial_c^\parallel+n_c\partial^\perp$. Integrating the parallel derivative components by parts we obtain

\be
G_{ef}^{abcd}\int_{\partial M^+}\left[(\overline{p}_{ab}n_cn_d)\partial^\perp\phi^{ef}-(\overline{p}_{ab,c}n_d+\partial_d^\parallel(\overline{p}_{ab}n_c))\phi^{ef}\right]\,dS + (q\mbox{-term})=0
\ee

Note that we can choose $\phi^{ef}\mid_\Sigma$ and $\partial^\perp \phi^{ef}\mid_\Sigma$ independently hence we obtain two junction conditions satisfied on the surface $\Sigma$, given by

\begin{eqnarray}
G_{ef}^{abcd}\overline{\omega}_{ab}n_cn_d &=& 0 \label{eqn:metricdis}\\
G_{ef}^{abcd}((\overline{\omega}_{ab}n_c)_{,d} + \overline{\omega}_{ab,c}n_d - \partial^\perp(\overline{\omega}_{ab}n_c)n_d)&=& 0 \label{eqn:curvdis},
\end{eqnarray}

\noindent where we have defined
$\overline{\omega}_{ab}=\overline{p}_{ab}\!\mid_\Sigma-\overline{q}_{ab}\!\mid_\Sigma$
and derivatives as
$\overline{\omega}_{ab,c}=\overline{p}_{ab,c}\!\mid_\Sigma-\overline{q}_{ab,c}\!\mid_\Sigma$.
To interpret these conditions we look at two important cases.

\subsection{$\Sigma$ Spacelike}

Junction conditions at spacelike surfaces in general relativity are
well known and were given by Israel \cite{israel}; we show here that the above method reproduces these results.

We work in coordinates $(x^0,x^1,x^2,x^3)$ with $\eta_{ab}=\mbox{diag}(1,-1,-1,-1)$ and take $\Sigma$ to be given by $x^0=0$, hence $n_c=(1,0,0,0)$. Equation (\ref{eqn:metricdis}) can be split in $(0,0)$,$(0,i)$ and $(i,j)$ components, where $i,j,...=1,2,3$. The $(0,0)$ and $(0,i)$ components are automatically satisfied, and the $(i,j)$ components give the condition

\be\label{eqn:metricmatch}
\omega_{ij}=0,
\ee

\noindent hence the induced metrics on $\Sigma$ must agree. Note that
there is no restriction on the $(0,0)$ and $(0,i)$ components of the
discontinuity. This is related to the discontinuity in the normal to
the hypersurface, mentioned earlier. A direct calculation reveals that
the normal takes the form $n^\pm = (1-\frac{1}{2}\varepsilon
h^\pm_{00},\varepsilon h^\pm_{0i})$ on each side of the
hypersurface. Thus the discontinuity in the normal direction is
equivalent to the discontinuity in the $(0,0)$ and $(0,i)$ components
of the metric. Note also that this discontinuity could be transformed
away with a continuous (yet not $C^1$) coordinate transformation,
though such a transformation will change the differentiable structure
of the manifold.

Equation (\ref{eqn:curvdis}) can also be split, with the $(i,j)$ components giving the only condition, which we can write as

\be
\omega_{ij,0}-\omega_{0i,j}-\omega_{0j,i}-\eta_{ij}(\omega^{k}{}_{k,0}-2\omega^k{}_{0,k})=0.
\ee

\noindent which is simply the linearized form of the Israel matching
condition in the vacuum,

\be\label{eqn:israel}
\sembrack{K_{ij}-\eta_{ij}K}=0.  
\ee

\noindent Hence the extrinsic curvatures, $K_{ij}$, of $\Sigma$ must agree from both sides.

This result is then completely analogous to that in area Regge
calculus where we found that we could not have a discontinuity in the
metric, and also the extrinsic curvature had to match due to the zero
deficit angle condition.

\subsection{$\Sigma$ Null}

Junction conditions for null hypersurfaces have been studied in
\cite{clarkedray:null,taub}; however all assume the induced metrics on the null
hypersurface agree. Following on from \cite{barrett:refract}, and
since we are only considering the linearized case, we do not make such an assumption, and see what the field equations demand.

We work in coordinates $(u,v,y,z)$ with metric $ds^2=2du\,dv-dy^2-dz^2$. Let $\Sigma$ be given by $u=0$ hence $n_c=(1,0,0,0)$, but note $\partial^\perp=\frac{\partial}{\partial u}=l^c\partial_c$, where $l_c=(0,1,0,0)$. Despite this the above integrations by parts can still be done, giving the same equations (\ref{eqn:metricdis}, \ref{eqn:curvdis}).

Let indices $\alpha,\beta,...$ represent the spatial $(y,z)$ directions. Solving equation (\ref{eqn:metricdis}), the only non-trivial solutions are the $(u,u)$, $(u,\alpha)$ and $(\alpha,\beta)$ components, which can be combined into the following one equation

\be\label{eqn:areamatch}
\overline{\omega}_{ab}n^b=0.
\ee

\noindent This is exactly the condition obtained by Barrett
\cite{barrett:refract} which states that the areas of spacelike
2-surfaces is continuous across $\Sigma$. This is therefore the
linearized metric of a \emph{refractive wave} spacetime. Solving for $\omega_{ab}$, we can see explicitly the form of the metric discontinuity.

\be\label{eqn:areamatchmatrix}
\omega_{ab}=\left( \begin{array}{cccc}
\omega_{uu} & \omega_{uv} & \omega_{uy} & \omega_{uz} \\
\omega_{uv} & 0 & 0 & 0 \\
\omega_{uy} & 0 & \omega_{yy} & \omega_{yz} \\
\omega_{uz} & 0 & \omega_{yz} & -\omega_{yy} \end{array} \right) 
\ee

\noindent Note again that the $\omega_{ua}$ terms can be transformed
away with a continuous transformation, yet the $\omega_{\alpha\beta}$
terms cannot.

This result is also completely analogous to that obtained in area
Regge calculus, where we found a discontinuity which, in the
linearized situation of Equation (\ref{eqn:linrefract}), is precisely of this form.

Solving equation (\ref{eqn:curvdis}) and using equation (\ref{eqn:areamatch}) to simplify the results we obtain four conditions, which we write in the following suggestive forms

\begin{eqnarray}
\eta^{\alpha\beta}(\omega_{\alpha u,\beta}+\omega_{\beta u,\alpha}-\omega_{\alpha\beta,u})&=&0, \label{eqn:nullcurv1}\\
\omega^\beta{}_{\alpha,\beta}+(\omega_{v\alpha,u}-\omega_{vu,\alpha}-\omega_{u\alpha,v})&=&0, \label{eqn:nullcurv2}\\
\omega_{uv,v}+\omega_{vu,v}-\omega_{vv,u}&=&0, \label{eqn:nullcurv3}\\
\omega_{v\alpha,\beta}+\omega_{v\beta,\alpha}-\omega_{\alpha\beta,v}&=&0. \label{eqn:nullcurv4}
\end{eqnarray}

Now, for a null hypersurface we can define various fundamental forms,
see \cite{clarkedray:null} for more details. These are the
\emph{internal} second fundamental form $\chi_{\alpha \beta}$, the
\emph{external} second fundamental form $\psi_{\alpha \beta}$, the
\emph{normal} fundamental form $\eta_\alpha$, and a quantity
$\omega$. In this case these are given by

\begin{eqnarray}
\chi_{\alpha\beta}&=&\nabla_\alpha n_\beta,\\
\psi_{\alpha\beta}&=&\nabla_\alpha l_\beta,\\
\eta_{\alpha}&=&\nabla_\alpha l_v,\\
\omega&=&\nabla_v l_v,
\end{eqnarray}

\noindent which we can calculate for this linearized situation, and hence rewrite equations (\ref{eqn:nullcurv1}-\ref{eqn:nullcurv4}) in terms of these quantities

\begin{eqnarray}
\sembrack{\eta^{\alpha\beta}\psi_{\alpha\beta}}&=&0,\\
\sembrack{\Gamma_{\alpha\beta}{}^\beta+2\eta_\alpha}&=&0,\\
\sembrack{\omega}&=&0,\\
\sembrack{\chi_{\alpha\beta}}&=&0.
\end{eqnarray}

\noindent The quantity $\Gamma_{\alpha\beta}{}^\beta$
in the second of these equations does not appear to be interpretable in terms of
our fundamental forms.

In the case where the induced metric is assumed to be continuous these
conditions reduce to those obtained by Clarke and Dray
\cite{clarkedray:null}, namely that
$\sembrack{\eta^{\alpha\beta}\psi_{\alpha\beta}}=\sembrack{\eta_\alpha}=\sembrack{\omega}=0$.
However, here we have the more general case allowing for the
discontinuous metrics present in the linearized thoery.

\section{Conclusions and Outlook}

We have shown how to construct non-trivial solutions to area Regge
calculus. In particular we have shown the existence of solutions with
a discontinuity in the metric along a null hypersurface. These
solutions are precisely the refractive wave spacetimes introduced by
Barrett \cite{barrett:refract}.

The relation between our results for area Regge calculus and those of the
linearized Einstein equations is also apparent. At spacelike
hypersurfaces both theories predict a matching of the induced metrics,
and of the extrinsic curvature. At null hypersurfaces, both theories
predict the discontinuous metric of a refractive wave spacetime. We
would also like to be able to compare the curvature matching
conditions in this case.

This comparison with linearized gravity fits in nicely with the ideas
of Rocek and Williams \cite{discrete}. They calculate the weak field expansion
of area Regge calculus and show the dynamics of the theory to be
equivalent to that of edge length variable Regge calculus. Since at
the perturbative level, in the long wavelength limit, Regge
calculus is equivalent to Einstein's theory \cite{qrc,qofrc}, so also
will area Regge calculus.

Despite this perturbative/dynamical relation, area Regge calculus is
clearly not so simply associated with the full theory of general
relativity. The refractive wave solution of area Regge calculus is not
a solution of the full non-linear theory of relativity, since for such
a metric the curvature tensor will not be well defined.

The next step in the study of area Regge calculus would be to look for
more complex and interesting solutions. It appears to be
straightforward to combine refractive wave solutions so long as they
are moving parallel or antiparallel, the result being a spacetime with
multiple discontinuities. In double null coordinates $(u,v,y,z)$ the
discontinuities are along surfaces of constant $u$, and surfaces of
constant $v$. These discontinuities will thus intersect, and we can
check that the field equations are still satisfied where this
occurs. We can also consider geodesics which start parallel, but
travel either side of such an intersection to a new region of
space. Here the geodesics will no longer be necessarily parallel, which
reveals the non trivial nature of these solutions. 

More general solutions to area Regge calculus can presumably be
found. Whether we can relate the properties of these solutions, such
as geodesic behaviour, to properties of solutions to general
relativity remains to be seen.

The ultimate goal will be to relate area Regge calculus back to the
quantum theories of gravity from which it was born, in order to
understand to a greater extent what these theories are telling us
about the nature of space, time and gravitation.

\section*{Acknowledgements}

We would like to thank John Barrett and John Stewart for helpful
conversations. The work was supported in part by the UK Engineering
and Physical Sciences Research Council and the UK Particle Physics and
Astronomy Research Council.

\appendix
\section{Appendix}

\subsection{Rank of $G$}\label{sec:rankg}

We show here the matrix $G$ of Equation (\ref{eqn:mat1}) has rank 3. First we can define a vector
$\boldsymbol{s}_{AB}$ by

\be
\boldsymbol{s}^T_{AB}=\left(
\left|\begin{array}{cc} Y_A & Z_A \\ Y_B & Z_B \end{array}\right|^2,
\left|\begin{array}{cc} Z_A & X_A \\ Z_B & X_B \end{array}\right|^2,
\left|\begin{array}{cc} X_A & Y_A \\ X_B & Y_B \end{array}\right|^2\right).
\ee

\noindent Then $G$ takes the form

\be
G^T=\left(\boldsymbol{s}_{13}, \boldsymbol{s}_{15},
\boldsymbol{s}_{17}, \boldsymbol{s}_{26}, \boldsymbol{s}_{27},
\boldsymbol{s}_{37} \right).
\ee

\noindent Now we know the coordinates of vertices 3, 5, 6 and 7 are
given in terms of those of vertices 1, 2 and 4. So, using the
properties of determinants we have various relations to simplify the
matrix. We can then use column operations on the matrix $G^T$ to put it in the form

\be
G^T = \left(\boldsymbol{s}_{24}, \boldsymbol{s}_{41},
\boldsymbol{s}_{12}, \ast, \ast, \ast \right).
\ee

\noindent Without loss of generality we choose the coordinates of our
lattice basis vector as $\boldsymbol{x}_1^T=(a,b,c)$,
$\boldsymbol{x}_2^T=(0,d,e)$ and $\boldsymbol{x}_4^T=(0,0,f)$ where $a,d,f\ne 0$. The matrix $G$ will then take the form

\be
G=\left(\begin{array}{ccc}
d^2f^2 & 0 & 0 \\
\ast & a^2f^2 & 0 \\
\ast & \ast & a^2d^2 \\
\ast & \ast & \ast \\
\ast & \ast & \ast \\
\ast & \ast & \ast 
\end{array}\right),
\ee

\noindent thus, since $a,d,f\ne 0$, this matrix will have rank 3.

\subsection{Rank of $\tilde G$}\label{sec:ranktildeg}

We show here that the matrix $\tilde G$ of Equation (\ref{eqn:nulllatvar}) has rank 3. First define the vector $\boldsymbol{r}_{AB}$ by

\be
\boldsymbol{r}^T_{AB}=\left(
\left|\begin{array}{cc} Y_A & Z_A \\ Y_B & Z_B \end{array}\right|,
\left|\begin{array}{cc} Z_A & V_A \\ Z_B & V_B \end{array}\right|,
\left|\begin{array}{cc} V_A & Y_A \\ V_B & Y_B \end{array}\right|\right).
\ee

\noindent Then we can write $\tilde G$ in the form

\be
\tilde G^T = \left(\boldsymbol{r}_{13}, \boldsymbol{r}_{15},
\boldsymbol{r}_{17}, \boldsymbol{r}_{26}, \boldsymbol{r}_{27},
\boldsymbol{r}_{37}\right).
\ee

\noindent Now we know that the coordinates of the vertices 3, 5, 6 and 7 can be given in terms of those of vertices 1, 2 and 4. So, using the properties of determinants we have for example $\boldsymbol{r}_{13}=\boldsymbol{r}_{11}+\boldsymbol{r}_{12}=\boldsymbol{r}_{12}$, and likewise for the other entries of $\tilde G$. Thus we can use column operations to put $\tilde G^T$ in the form

\be
\tilde G^T = \left(\boldsymbol{r}_{24}, \boldsymbol{r}_{41},
\boldsymbol{r}_{12}, 0, 0, 0\right).
\ee

\noindent The first three rows of $G$ are then just the cofactor matrix of 

\be
S=\left(\begin{array}{ccc}
V_1 & Y_1 & Z_1 \\
V_2 & Y_2 & Z_2 \\
V_4 & Y_4 & Z_4
\end{array}\right),
\ee

\noindent and since the three vectors along the edges 1, 2 and 4 are linearly independent we have $\mbox{det}\,S \ne 0$. Hence, the cofactor matrix of $S$ will also be invertible, which implies that $\tilde G$ will have rank 3.

\bibliographystyle{unsrt}

\begin{thebibliography}{10}

\bibitem{ponzanoregge}
G.~Ponzano and T.~Regge.
\newblock Semiclassical limit of Racah coefficients.
\newblock {\em Spectroscopic and group theoretical methods in
  physics}, ed F. Bloch et al. (Amsterdam: North-Holland), pages 1--58, 1968.


\bibitem{regge:orig}
Tulio Regge.
\newblock General relativity without coordinates.
\newblock {\em Nuovo Cimento} \textbf{19}: 558--571, 1961.

\bibitem{discrete}
Tulio Regge and Ruth~M. Williams.
\newblock Discrete structures in gravity.
\newblock {\em J. Math. Phys.} \textbf{41}: 3964--3984, 2000.

\bibitem{barrettcrane}
John~W. Barrett and Louis Crane.
\newblock Relativistic spin networks and quantum gravity.
\newblock {\em J. Math. Phys.} \textbf{39}: 3296--3302, 1998.

\bibitem{rovelli}
Carlo Rovelli.
\newblock Basis of the Ponzano-Regge-Turaev-Viro-Ooguri quantum-gravity model
  is the loop representation basis.
\newblock {\em Phys. Rev.} D \textbf{48}: 2702--2707, 1993.

\bibitem{noteonareavar}
John~W. Barrett, Martin Rocek, and Ruth~M. Williams.
\newblock A note on area variables in Regge calculus.
\newblock {\em Class. Quantum Grav.} \textbf{16}: 1373--1376, 1997.

\bibitem{qrc}
M.~Rocek and Ruth~M. Williams.
\newblock Quantum Regge calculus.
\newblock {\em Phys. Lett.} \textbf{104B}: 31--37, 1981.

\bibitem{qofrc}
M.~Rocek and Ruth~M. Williams.
\newblock The quantization of Regge calculus.
\newblock {\em Z. Phys.} C \textbf{21}: 371--381, 1984.

\bibitem{barrett:refract}
John~W. Barrett.
\newblock Refractive gravitational waves.
\newblock {\em gr-qc/0011051}, Nov 2000.

\bibitem{hartley}
David~Hartley et~al.
\newblock Tensor distributions on signature-changing space-times.
\newblock {\em Gen. Rel. Grav.} \textbf{32}: 491--503, 2000.

\bibitem{dray}
Tevian Dray.
\newblock Tensor distributions in the presence of degenerate metrics.
\newblock {\em Int. J. Mod. Phys.} \textbf{D6}: 717--740, 1997.

\bibitem{geroch}
Robert Geroch and Jennie Traschen.
\newblock Strings and other distributional sources in general relativity.
\newblock {\em Phys. Rev.} D \textbf{36}: 1017--1031, 1987.

\bibitem{penrose}
R. Penrose.
\newblock The geometry of impulsive gravitational waves.
\newblock {\em General Relativity. Papers in honour of J. L. Synge}
ed. L. O'Raifeartaigh. (Clarendon, Oxford), pages 101--115, 1972.

\bibitem{israel}
W.~Israel.
\newblock Singular hypersurfaces and thin shells in general relativity.
\newblock {\em Nuovo Cimento} B \textbf{44}: 1, 1966.

\bibitem{clarkedray:null}
C.~J.~S. Clarke and Tevian Dray.
\newblock Junction conditions for null hypersurfaces.
\newblock {\em Class. Quantum Grav.} \textbf{4}: 265--275, 1986.

\bibitem{taub}
A.~H. Taub.
\newblock Space-times with distribution valued curvature tensors.
\newblock {\em J. Math. Phys.} \textbf{21}: 1423--1431, 1980.


\end{thebibliography}

\end{document}